\documentclass[aps,prl,a4paper,twocolumn,superscriptaddress,epsf,showpacs,showemail]{revtex4}

\usepackage{graphicx}
\usepackage{dcolumn}
\usepackage{bm}
\usepackage{amsmath}
\usepackage[english]{babel}
\usepackage[utf8]{inputenc}

\begin{document}


\title{Sagnac interferometry with a single atomic clock}



\author{R.~Stevenson}
\author{M.~Hush}
\author{T.~Bishop}
\author{I.~Lesanovsky}
\author{T.~Fernholz}
\affiliation{School of Physics \& Astronomy, University of Nottingham, Nottingham NG7 2RD, United Kingdom}
\email{thomas.fernholz@nottingham.ac.uk}


\date{\today}

\begin{abstract}
We theoretically discuss an implementation of a Sagnac interferometer with cold atoms. In contrast to currently existing schemes our protocol does not rely on any free propagation of atoms. Instead it is based on superpositions of fully confined atoms and state-dependent transport along a closed path. Using Ramsey sequences for an atomic clock, the accumulated Sagnac phase is encoded in the resulting population imbalance between two internal (clock) states. Using minimal models for the above protocol we analytically quantify limitations arising from atomic dynamics and finite temperature. We discuss an actual implementation of the interferometer with adiabatic radio-frequency potentials that is inherently robust against common mode noise as well as phase noise from the reference oscillator.
\end{abstract}

\pacs{07.60.Ly, 03.75.-b, 03.75.Dg}

\keywords{atom traps, matter-wave interferometry}

\maketitle



The Sagnac effect enables interferometric measurements of rotation with high precision \cite{Schreiber:13}. For example, the large Wettzell laser gyroscope achieves a theoretical resolution of $10^{-11}\textrm{rad}/\sqrt{\textrm{s}}$ \cite{Schreiber:11}. Interferometers based on matterwaves instead of light promise resolution enhancement by orders of magnitude that scales with particle mass \cite{Clauser:88}, see~\cite{Barrett:14} for a recent review. Despite the immense challenges in achieving similar particle flux and interferometer areas as with photons, atomic gyroscopes \cite{Lenef:97,Gustavson:97} have reached performance levels that should enable applications in fundamental physics, geodesy, seismology, or inertial navigation. Atom interferometers \cite{Cronin:09} have been demonstrated with record sensitivities below $10^{-9} \textrm{rad}/\sqrt{\textrm{s}}$ \cite{Gustavson:00, Durfee:06} outperforming commercial navigation sensors by orders of magnitude. Recent experiments aim at geodetic \cite{Stockton:11} and navigational applications combining multi-axis measurements of acceleration and rotation \cite{Canuel:06, Dickerson:13}. Since free falling atoms require large apparatus size, ring shaped traps and guided interferometers have been proposed \cite{Morizot:06, Lesanovsky:07, Japha:07, Baker:09} and demonstrated \cite{Sauer:01, Jo:07, Wu:07, Ryu:07, Zawazki:10, Marti:15} for a variety of geometries and levels of sophistication, e.g., using soliton dynamics in Bose-Einstein condensates to enhance sensitivity by non-linear interactions or to prevent wavepacket dispersion \cite{Veretenov:07, McDonald:14, Helm:15}.

So far, the paradigm for matter wave Sagnac interferometry relies on DeBroglie waves and thus on free propagation of atoms either in free fall or within waveguides. However, the Sagnac effect can be expressed as a propertime difference experienced by two observers moving in opposite directions along closed paths and has indeed been measured with atomic clocks flown around Earth \cite{Hafele:72}. Inspired by this, we investigate an interferometer comprised of a single atomic clock. It uses the acquired phase shift between atoms in two different internal clock (spin) states that are each fully confined in atom traps but separately displaced. This approach offers a high degree of control over atomic motion, removing velocity dependent effects and phase front and interferometric stability requirements of laser beams. It improves control of heating from waveguide corrugations and avoids wave packet dispersion allowing for multiple revolutions. We discuss a robust Ramsey interferometer scheme, formulate quantum mechanical models for different trap dimensionalities, and investigate effects that arise from motional excitation and non-adiabatic transport.

\begin{figure}
\centering
\includegraphics[width=0.4\textwidth]{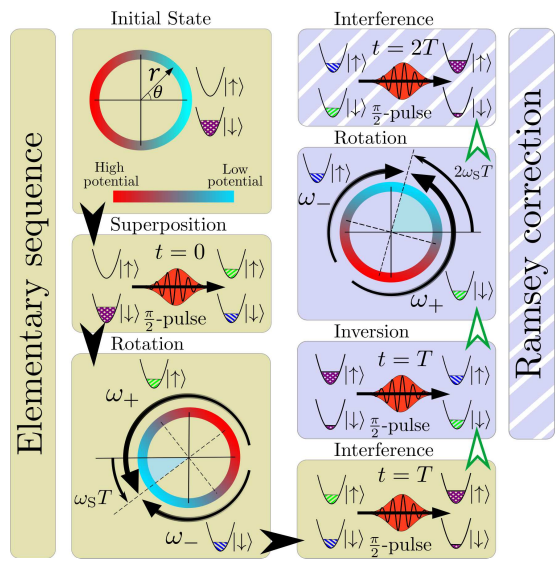}
\caption{\label{fig:ExperimentalProtocol} Experimental sequence depicted in an inertial frame. Starting with atoms prepared in $|{\downarrow}\rangle$ located at $\theta = 0$, a $\pi/2$-pulse generates a superposition of two non-degenerate internal states. Atoms in $|({ \uparrow}){ \downarrow}\rangle$ are then transported along a circular path in (anti-)clockwise direction. After a half-revolution, a second pulse converts any phase shift into population difference, which is measured in the elementary sequence (black arrows). An extended Ramsey sequence (green arrows) can be used to achieve full common path operation. Here, the $\pi/2$-pulse at time $T$ is extended to a $\pi$-pulse, fully inverting the atomic states. Transport is continued such that all atoms complete a full revolution before converting and measuring the phase difference at time $2T$.}
\end{figure}
The Sagnac effect for fully confined atoms can be depicted in an inertial frame, as seen in Fig.~\ref{fig:ExperimentalProtocol}. Two independent traps, each containing atoms of rest mass $m$ are displaced around a ring of radius $r$. Starting from a common angular position at $\theta=0$, the traps are moved along counter-propagating trajectories and recombined at multiples of the half-revolution time $T$. From the experimenter's point of view, who defines the trajectories, this happens on the opposite side of the ring and at the original starting point. But if the laboratory frame is rotating at angular frequency $\omega_\text{S}$, the first recombination will occur at $\theta=\pi+\omega_\text{S} T$. In the inertial frame, the traps will therefore be displaced at different average angular speeds $\omega_{\pm}=\pi/T\pm\omega_\text{S}$, leading to a propertime difference of $\Delta\tau_\text{p}\approx 2\pi\omega_\text{S} r^2/c^2$ for the two co-moving rest frames, proportional to interferometer area. In these co-moving frames each atomic state can be described as evolving at its respective Compton frequency $\omega_C=mc^2/\hbar$ \cite{Lan:13, Peil:14}, leading to a phase difference
$\Delta(\omega_\text{C}\tau_\text{p})\approx\omega_\text{C}\Delta\tau_\text{p}+\Delta\omega_\text{C}\tau_\text{p}$ for non-relativistic speeds and energies \footnote[1]{Apart from evolution due to the confining potentials, which we endeavour to make the same in both paths.}. The first term, which is equivalent to the propagation phase difference in the inertial frame, leads to the Sagnac phase for a half revolution $\phi_\text{S}\approx 2\pi\omega_\text{S} r^2m/\hbar$, which advances any dynamical phase $\Delta\omega_\text{C}\tau_\text{p}\approx \Delta ET/\hbar$ resulting from energy differences $\Delta E$ that can be included in the rest mass, e.g., different internal energies of two clock states. This argument shows that the Sagnac phase can indeed be measured accurately in a fully guided setting, as long as the internal energy difference is precisely known or compensated, and shifts due to confinement or external effects remain identical when observed in the two rest frames. The following will discuss a specific implementation with cold atoms.

\emph{Interferometer sequence.} Our scheme requires state-dependently controlled trapping potentials moved around a ring in combination with an interferometer sequence used for atomic clocks, as shown in Fig.~\ref{fig:ExperimentalProtocol}. In the elementary sequence we study here, each atom is initially trapped at $\theta=0$ and prepared in a superposition of two non-degenerate spin states $|\Psi\rangle=\frac{1}{\sqrt{2}}(|{\downarrow}\rangle+|{ \uparrow}\rangle)$ by starting in $|{\downarrow}\rangle$ and driving a resonant $\pi/2$-pulse derived from a stable reference clock. The state-dependency is then used to move atoms in state $|({ \uparrow}){ \downarrow}\rangle$ (anti-)clockwise around the ring. When the two components recombine on the opposite side, they will have acquired a relative phase difference, which is measured by driving a second $\pi/2$-pulse with an adjustable phase $\phi_\text{ref}$, converting the phase difference into population difference.

In order to remove constant perturbations of the two spin state energies or, equivalently, a constant detuning of the reference clock, a spin echo sequence can be used. The $\pi/2$-pulse at time $T$ is extended to a $\pi$-pulse, exchanging states $|{\downarrow}\rangle$ and $|{ \uparrow}\rangle$, before rotating the state dependent traps in the opposite direction such that each component completes a full revolution over time $2T$. As before, a final $\pi/2$-pulse converts phase difference into measurable number difference. This sequence removes the time-dependent dynamical phase, because all atoms spend half the observer's time in each spin state. While this prevents operation as an atomic clock, it does, however, not remove the path dependent Sagnac phase. This procedure also cancels effects from constant but spatially dependent energy shifts as all atoms travel the same paths in the same spin states. Due to the common path for a full revolution, dynamical phases caused by constant external acceleration, gravitation or other static potentials do not affect the measurement.

\emph{Guided interferometer models.} In the following, we analyze the effects of fully confined transport and determine conditions that allow for reliable measurements of the Sagnac phase. We neglect any interactions or mixing of the two spin states and describe the dynamics of the interferometer by a Hamiltonian of the form $\hat H = \hbar \omega [\hat H_{ \uparrow}|{\uparrow}\rangle \langle{ \uparrow}| + \hat H_{ \downarrow}|{\downarrow}\rangle\langle{ \downarrow}|]$.  Furthermore, we assume identical shapes for the two state-dependent potentials and equal and opposite paths in the laboratory frame. For the elementary sequence shown in Fig.~\ref{fig:ExperimentalProtocol} and atoms starting in motional ground state $|g\rangle$, the final atomic state can be expressed using unitary evolution operators $|\Psi(T)\rangle = \hat P(\phi_\text{ref}) \hat U(T) \hat P(0) |g\rangle\otimes|{ \downarrow}\rangle$, where $\hat U(T) = \hat U_{ \uparrow}(T)\otimes \hat U_{ \downarrow}(T)$ is the evolution imposed by the Hamiltonian and $\hat P(\phi)$ describes a $\pi/2$-pulse with phase $\phi$. The measured signal is the population difference $\langle \hat\sigma_\text{z} \rangle$, where $\hat\sigma_\text{z}=|{ \downarrow}\rangle\langle {\downarrow}|-|{ \uparrow}\rangle\langle{ \uparrow}|$. This expression simplifies to:
\begin{align}
\label{eqn:MatrixElement} \langle\hat\sigma_z\rangle = \frac{1}{2}\langle g | \hat U^\dagger_{ \downarrow}(T) \hat U_{ \uparrow}(T) |g\rangle e^{i \phi_\text{ref}} + \text{H.c.},
\end{align}
Control of $\phi_\text{ref}$ enables interferometer operation near maximal dependence on laboratory rotation. Accordingly, we define the (dimensionless) sensitivity or scale factor as:
\begin{align}
\Sigma = \underset{\phi_\text{ref}}{\text{max}}\left|\frac{d\langle\hat\sigma_\text{z}\rangle}{d \omega_\text{S}}\right|\omega.
\label{eqn:Sensitivity}
\end{align}

\emph{One-dimensional model.} First, we consider an idealised situation where the atoms are tightly confined to a ring of radius $r$, thus restricting the motional degrees of freedom to the azimuthal coordinate $\theta$. Within this ring we assume that two harmonic potentials with trapping frequency $\omega$ are displaced by the experimenter in opposite directions at angular speed $\omega_\text{P}(t)$. In the laboratory frame, both paths end on the opposite side of the ring at $t=T$, imposing the condition $\int_0^T \omega_\text{P}(t) dt = \pi$. Transforming the Hamiltonian to a state-dependent rotating frame that keeps both potentials stationary leads to:
\begin{align}
\hat H_{\uparrow(\downarrow)} &= \hat a^\dagger \hat a + i\frac{R}{\sqrt 2}\left[\Omega_\text{S} + \eta_{\uparrow(\downarrow)} \Omega_\text{P}(\tau)\right] (\hat a -\hat a^\dagger),
\label{eqn:1DHam}
\end{align}
where we introduced $\eta_{\uparrow(\downarrow)} = + (-) 1$ and dimensionless parameters $\tau = \omega t$, $\Omega_\text{S} = \omega_\text{S}/\omega$, $\Omega_\text{P}(\tau) = \omega_\text{P}(\tau)/\omega$ and $R = r/x_\mathrm{ho}$ with $x_\mathrm{ho}=\sqrt{\hbar/m\omega}$ being the harmonic oscillator length. The Hamiltonians in Eq.~(\ref{eqn:1DHam}) describe forced harmonic oscillators whose unitary time-evolution operators can be expressed in the form of displacement operators $\hat U_{\uparrow(\downarrow)}=\exp(\alpha_{\uparrow(\downarrow)}^* \hat a - \alpha_{\uparrow(\downarrow)} \hat a^\dagger) \exp(i\phi_{\uparrow(\downarrow)})$ via the Magnus expansion \cite{Blanes:09}. We do not provide here the rather lengthy explicit expressions of displacement $\alpha_{\uparrow(\downarrow)}$ and phase $\phi_{\uparrow(\downarrow)}$. Substituting the evolution operators into Eq.~(\ref{eqn:MatrixElement}) yields the population difference after the interference step at time $T$ (see Fig. \ref{fig:ExperimentalProtocol}):
\begin{align}
\label{eqn:Exact}
\langle\hat\sigma_\text{z}\rangle &=C_\text{1D}\cos \left(\phi_\text{S}+\phi_\text{ref}\right),
\end{align}
which consists of two factors. The first one is the contrast $C_\text{1D} = e^{-|\Delta \alpha|^2/2}$, which depends only on the final relative coherent displacement of the two spin components $\Delta\alpha=\alpha_{ \uparrow} - \alpha_{ \downarrow} = \sqrt 2 r/x_\text{ho} \int_0^{T} \omega_\text{P}(t)e^{-i\omega t}dt$. The second factor is the oscillatory part of the signal which indeed depends on the Sagnac phase $\phi_\text{S} =2\pi m r^2 \omega_\text{S}/\hbar$.

This result shows that the Sagnac phase difference accumulated by the atoms remains independent of the temporal profile $\omega_\mathrm{P}(\tau)$ of the path taken. However, the interferometer contrast, and therefore the signal's sensitivity to rotation is reduced if the final states of the two components are no longer in the ground state of the trap but (symmetrically) displaced. Any choice of temporal path that does not contain Fourier components at the trapping frequency, i.e., for which $\int_0^{T} \omega_\text{P}(t)e^{-i\omega t}dt=0$, will achieve maximum contrast by ensuring that the two wavepackets overlap completely and appear stationary, i.e. $\Delta\alpha=0$. The maximum speed at which this can be achieved is in principle only limited by the maximum potential energy at which the harmonic oscillator approximation for the confining potentials remains valid.

\emph{Two-dimensional model.} The one-dimensional model is oversimplified due to the assumption of an infinitely strong radial confinement. In any practical implementation non-negligible radial forces will occur which depend on the rotational speed and which are, in particular, different for the two spin states when $\Omega_\text{S}\neq 0$. To understand how these inevitable effects impact on the operation of the Sagnac interferometer we consider an exactly solvable two-dimensional model in which atoms are held in the isotropic and harmonic oscillator potential:
\begin{equation}
V(x,y) = \frac{1}{2}m\omega^2\left[(x-\cos\hat\theta(t))^2 + (y-\sin\hat\theta(t))^2\right].
\end{equation}
As in the one-dimensional example, both spin components travel in opposite directions. Spin-dependent trap motion is introduced using $\hat\theta(t)= \int_0^t du [\omega_\text{S} + \hat \sigma_\text{z}\omega_\text{P}(u)]$. It turns out that a particularly simple analytical description of the system is achieved by introducing the operators $\hat A_\pm = \frac{1}{2 x_\text{ho}}(\pm i \hat x + \hat y) + \frac{x_\text{ho}}{2}(\pm i\frac{d}{dx} + \frac{d}{dy})$. The Hamiltonian is then given by: \begin{align}
\hat H_{\uparrow(\downarrow)} &= \hat H_{+,\uparrow(\downarrow)} + \hat H_{-,\uparrow(\downarrow)}\\
\hat H_{\pm,\uparrow(\downarrow)} &= [1 \pm \Omega_\text{S}  \pm  \eta_{\uparrow(\downarrow)}
\Omega_\text{P}(\tau)]\hat A_\pm^\dagger \hat A_\pm \mp
\hbar R \frac{\hat A_\pm-\hat A_\pm^\dagger}{2 i},\nonumber
\end{align}
where we used the same dimensionless quantities as in Eq.~(\ref{eqn:1DHam}). After transforming into an interaction picture using the transformation $\hat W = \hat W_+ \, \hat W_-$ with $\hat W_\pm = e^{\mp \hat \sigma_\text{z} i \int_0^{\omega T} d\tau\Omega_\text{P}(\tau) \hat A_\pm^\dagger  \hat A_\pm}$ the problem separates into linearly forced harmonic oscillators. For the elementary sequence of the interferometer protocol we perform a half-rotation of the two traps in opposite directions. As before, this imposes the condition $\theta_\text{P}(\omega T) = \pi$ on the angular displacement of the potentials in the laboratory frame $\theta_\text{P}(\tau) = \int_0^\tau d\tau' \Omega_\text{P}(\tau')$. After the interference step at time $T$ (see Fig. \ref{fig:ExperimentalProtocol}) the interferometer signal is given by:
\begin{align}
\langle\hat\sigma_\text{z}\rangle = C_+ C_- \cos\left(\phi_+ - \phi_- + \phi_\text{ref}\right),\label{eqn:2Dcontrast}
\end{align}
which depends on the phases:
\begin{widetext}
\begin{align}
\phi_\pm &= \frac{ R^2}{1\pm\Omega_\text{S}} \int_0^{\omega T}d\tau \sin[\theta_\text{P} (\tau)] \sin[(1\pm\Omega_\text{S})\tau]-\frac{R^2}{4}\int_0^{\omega T} d\tau \int_0^{\omega T} d\tau' \sin\left[\theta_\text{P}(\tau')+\theta_\text{P}(\tau)+(1\pm\Omega_\text{S})(\tau'-\tau)\right]\nonumber\\
& -\frac{R^2}{2} \int_0^{\omega T} d\tau \int_0^{\tau} d\tau' \cos
\left[ (1\pm\Omega_\text{S})(\tau'-\tau)\right]\sin \left[
\theta_\text{P}(\tau')-\theta_\text{P}(\tau)\right]
\end{align}
\end{widetext}
and the contrast coefficients:
\begin{align}
C_\pm =  \exp\left[-\frac{R^2}{2} \left|\int_0^{\omega T} d\tau
\sin[\theta_\text{P}(\tau)] e^{i(1\pm \Omega_\text{S})\tau}
 \right|^2 \right].\label{eqn:Contrast2D}
\end{align}

\begin{figure}
\centering
\includegraphics[width=0.4\textwidth]{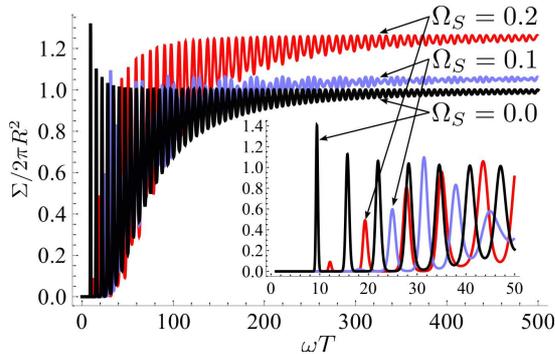}
\caption{\label{fig:Sensitivity} Sensitivity $\Sigma$ for the two-dimensional interferometer with $R=10$ and flat-top speed profile $\Omega_\text{P}(\tau) = \pi/\omega T$ for $0<\tau<\omega T$. Plots for $\Omega_\text{S} = 0, 0.1, 0.2$ are shown in black, blue and red. The inset shows the same data for $0<\omega T<50$.}
\end{figure}

As an example, we show the sensitivity of the two-dimensional interferometer in Fig.~\ref{fig:Sensitivity} for a ring of radius $R=10$ (harmonic oscillator lengths) and for different values of the rotational speed $\Omega_\text{S}$. The results were obtained for a temporal path of constant speed (flat-top profile), i.e., $\Omega_\text{P}(\tau) = \pi/\omega T$ for $0<\tau<\omega T$. For slow path speeds ($\omega T\gg 1$) the sensitivity approaches the adiabatic value $\Sigma_\mathrm{ad}=\frac{d}{d \Omega_\text{S}}2 \pi \Omega_\text{S} R^2 /(1-\Omega_\text{S}^2)^2$. For increasingly faster cycles non-adiabatic effects, i.e., the sloshing motion of the atomic wavepackets in the individual traps due to sudden acceleration, give rise to oscillations in the sensitivity. In the extreme case (non-overlapping wavepackets at time $T$) the sensitivity approaches zero. Conversely, times of maximum overlap result in peaked sensitivity and are found at the approximate times $\omega T_k = (2k + 1)\pi/(1+\Omega_\text{S})$ for integer $k \geq 1$. As shown in the inset of Fig.~\ref{fig:Sensitivity}, sensitivities at these times are close to or even larger than the adiabatic limit $\Sigma_\mathrm{ad}$ for small $\Omega_\text{S}$. As before, in principle this permits fast, i.e., non-adiabatic operation of the interferometer.

The data moreover shows that larger $\Omega_\text{S}$ as well as short operation times can result in higher sensitivity, caused by the interplay of three different effects. First, larger centrifugal forces lead to increased effective radius $R_\text{eff}$ and enclosed interferometer area. While this is the only effect in the adiabatic limit with $R_\text{eff}=R /(1-\Omega_\text{S}^2)$, it leads to non-linearly increasing sensitivity beyond the simple Sagnac effect due to rotation dependent area. Note, that for $|\Omega_\text{P}|+|\Omega_\text{S}|>1$ the centrifugal force overcomes the harmonic confinement and atoms become untrapped. Second, for non-zero $\Omega_\text{S}$, the two spin components experience different centrifugal forces and acquire
a phase difference from different potential energy in their respective traps, depending on their relative radial motion. Third, the interferometer contrast depends on the laboratory rotation. Overall, the most transparent situation is encountered at $\Omega_\text{S} = 0$, where the optimum phase reference angle is $\phi_\text{ref} = \pm \pi/2$ and the contrast coefficients are equal ($C_+ = C_-$). Here, similar but not identical to the one-dimensional case the contrast is maximised and independent of $\Omega_\text{S}$ by choosing a path such that $\int_0^{\omega T} d\tau \sin(\theta_\text{P}(\tau)) e^{i\tau}=0$.

\emph{Finite temperature.}
Finally, we consider interferometer operation with thermal states. We can use Glauber-Sudarshan distributions of the density matrix in terms of coherent states $\rho = \int d^2\epsilon p(\epsilon)|\epsilon\rangle\langle\epsilon|$ for each oscillator. For temperatures $\Theta$ well above the degeneracy temperature we find the distribution function $p(\epsilon) =\hbar\omega/(\pi k_\mathrm{B}\Theta) e^{-\hbar\omega|\epsilon-\alpha_\text{g}|^2/ k_\mathrm{B}\Theta}$ \cite{Gardiner:04}. A technical detail is the appearance of off-sets $\alpha_\mathrm{g}$ in the exponent. A state that is prepared in the laboratory frame appears displaced due to our definition of operators in the inertial frame. In the one-dimensional case we have $\alpha_\mathrm{g}=i R \Omega_\text{S}/\sqrt 2$, see Eq.~(\ref{eqn:1DHam}). Here, we obtain the thermal signal:
\begin{align}
\langle\hat\sigma_\text{z}\rangle_{\Theta} &=\int d^2\epsilon p(\epsilon) \langle \epsilon | \hat U^\dagger_{ \downarrow}(T) \hat U_{ \uparrow}(T) |\epsilon\rangle e^{i\phi_\text{ref}} + \text{H.c.}\nonumber\\
&=\langle\hat\sigma_\text{z}\rangle e^{-\frac{k_\mathrm{B}\Theta}{\hbar\omega}\left|\Delta\alpha\right|^2}=C_{\Theta}\cos(\phi_\text{S}+\phi_\text{ref}),
\end{align}
where $\langle\hat\sigma_\text{z} \rangle$ is the zero temperature result, see Eq.~(\ref{eqn:Exact}). The behaviour in the isotropic two-dimensional model is essentially identical but with the contrast dependent on the relative displacement in two dimensions.

This shows that finite temperatures result in unchanged interferometer signals, if motional excitation of the traps is avoided or cancelled after trap recombination ($\Delta\alpha = 0$). Otherwise, the zero-temperature reduction in contrast from imperfect state overlap is amplified. This is equivalent to a white light interferometer, where the required precision of wave packet overlap is given by the coherence length. E.g., for the one-dimensional case, a final relative displacement $\Delta x$ that is purely spatial, i.e., for equal momenta and $\Delta\alpha=\sqrt{m\omega/2\hbar}\langle\Delta \hat{x}\rangle$, the thermal contrast for high temperature can be expressed in terms of thermal wavelength $\lambda_{\Theta}=h/\sqrt{2\pi m k_\text{B} \Theta}$ and harmonic oscillator length $x_\mathrm{ho}$:
\begin{equation}
C_{\Theta}=e^{-\left(\frac{1}{2}+\frac{k_\mathrm{B}\Theta}{\hbar\omega}\right)\left|\Delta\alpha\right|^2} \approx e^{-\frac{\langle\Delta \hat{x}\rangle^2}{2}\left(\frac{1}{2 x_\mathrm{ho}^2}+\frac{\pi}{\lambda_{\Theta}^2}\right)}.
\end{equation}

\emph{Experimental implementation with dressed potentials.}
A suitable scheme to implement the required state-dependent transport of atomic clock states has been described recently \cite{Fernholz:07}, which uses radio-frequency (rf) fields to control atomic motion \cite{Lesanovsky:06, Hofferberth:06, Sherlock:11}. Interferometry with such rf potentials has already been demonstrated \cite{Schumm:05}. In the proposed scheme, circular waveguides are generated by using a cylindrically symmetric ring of zero magnetic field, produced by four coaxial current loops, and dressing it with rf fields of appropriate polarisation. We suggest to use magnetically trappable clock states of alkali atoms, e.g., $^{87}$rubidium \cite{Treutlein:04}, which have a nearly vanishing differential Zeeman shift and thus see almost identical ring potentials. State-dependence, which arises from equal but opposite g-factors of the two clock states, can be introduced in the azimuthal $\theta$-direction by a single, linearly polarised rf-field, whose direction of polarisation defines the recombinations points. Separation, guiding, and recombination of traps are simply driven by phase changes of this field, which maintains the symmetry of the traps and their transport apart from field imperfections. Additional robustness can be derived from the fact that the described potentials generate two stacked circular waveguides simultaneously, which differ in the direction of the underlying static field. This provides the opportunity to operate two closely spaced interferometers simultaneously, driven by the same rf-fields and reference clock but operating in opposite rotational senses. A differential measurement can then be used to reveal the Sagnac phase but remove phase noise from the required reference clock and symmetry breaking offsets.


\emph{Conclusion \& Outlook.} We have demonstrated that an atomic Sagnac interferometer can be implemented with fully confined atoms, at finite temperature, enabling new designs of compact devices. Beyond the principal effects discussed here, actual implementations will need to take into account and optimise effects resulting from interatomic collisions, corrugations and noise of trapping potentials, and interplay of thermal motion and finite length spin operations. Optimal control of atomic motion should allow for fast and robust interferometer operation which could in principle achieve sensitivity to rotation beyond the standard Sagnac effect.

\acknowledgments
We gratefully acknowlegde useful discussions with all members of
the Matterwave consortium. This research has been supported by the EU-FET Grant No. 601180 (MatterWave). IL acknowledges funding from the European Research Council under the European Union's Seventh Framework Programme (FP/2007-2013) / ERC Grant Agreement No. 335266 (ESCQUMA).


\end{document}